\input{aipcheck}

\documentclass[
    ,final            % use final for the camera ready runs
  ]
  {aipproc}

\layoutstyle{6x9}

\newcommand{\be}{\begin{equation}}
\newcommand{\ee}{\end{equation}}
\newcommand{\bse}{\begin{subequations}}
\newcommand{\ese}{\end{subequations}}
\newcommand{\bary}{\begin{eqnarray}}
\newcommand{\eary}{\end{eqnarray}}

\begin{document}

\title{On external shock model to explain the high-energy emission: GRB 940217, GRB 941017 and GRB 970217A }


\author{N. Fraija}{
  address={nifraija@astro.unam.mx}
}

\author{M. M. Gonz\'alez, J. L. Ramirez, R. Sacahui and W. H. Lee }{
  address={magda@astro.unam.mx, joselo.r.m@ciencias.unam.mx,  jsacahui@astro.unam.mx, wlee@astro.unam.mx }
}

%\author{W.H. Lee}{
 % address={}
 % ,altaddress={<author1 address>} % additional visiting address
%}

\begin{abstract}
 We present a leptonic model on the external shock context to describe the high-energy emission of GRB 940217,  GRB 941017  and GRB 970217A.   We argue that the emission consists of two components, one with a similar duration of the burst, and a second, longer-lasting GeV phase lasting hundred of seconds after the prompt phase.   Both components can be described as synchrotron self-Compton emission from a reverse and forward shock respectively. For the reverse shock, we analyze the synchrotron self-Compton in the thick-shell case. The calculated fluxes and break energies are all consistent with the observed values. 
\end{abstract}

\maketitle

%%%%%%%%%%%%%%%%%%%%%%%%%%%%%%%%%%%%%%%%%%%%
%% MAINMATTER
%%%%%%%%%%%%%%%%%%%%%%%%%%%%%%%%%%%%%%%%%%%%

\section{Introduction}

General hadronic and leptonic interpretations have been widely discussed to  explain photons with energies $\geq$ 100-MeV.   On hadronic models,  $\gamma$-ray radiation components  have been explained by  photo-hadronic interactions \citep{asa09, der00}.  On leptonic models,  high-energy  gamma-ray components  have been interpreted by Inverse Compton  \cite{pil98, pan00a}   and synchrotron self-Compton (SSC) \cite{sar01,wan01a,sac12,fra12a}.  In particular, Fraija et al. (2012) and Sacahui  et al. (2012) showed that  the short MeV and long-lasting GeV high-energy components  presented in GRB 980923 and GRB 090926A respectively, could come from  SSC emission in external shocks. In this work we apply the same model in GRB 940217, GRB 941017 and GRB 970417A  to describe their high-energy components, and by introducing standard values for the input parameters, we obtain break energies, fluxes, duration, etc in agreement with the  observed values. A brief description of the high-energy emission for the considered bursts is given following.
The first burst, GRB 940217 is one of the longest and also the most energetic burst. It was detected by the Compton Telescope (COMPTEL), the Energetic Gamma-Ray Experiment Telescope (EGRET) and Interplanetary Network (Ulysses/Burst and Transient Source Experiment, BATSE). The EGRET spark-chamber recorded 10 photons while the main emission was in progress.  Following this, an additional 18 photons were recorded for $\sim$ 5400 s, including an 18-GeV photon $\sim$ 4500 s after the main emission had ended. The total fluence  above 20 keV was (6.6$\pm$2.7)$\times$10$^{-4}$ erg cm$^{-2}$, as observed by BATSE large area detectors\cite{hur94}. 
The second burst, GRB 941017  was the first burst with clearly evidence of a high energy component different to the usual band function \citep{gon03}. Analysis of combined data from two detectors of the Compton observatory, BATSE's LAD and EGRET-TASC, was made to obtain the prompt spectra in an energy range between 30 keV to 200 MeV. This high energy component lasted longer than the burst $T_{90} = 77 s$ and it was described with a photon index of $\sim-1$ extending up to 200 MeV, and had a fluence above 30 keV of $6.5\times10^{-4}$ erg cm$^{-2}$, which is more than three times that estimated from the BATSE energy range alone. It did not exhibit an energy cut-off, suggesting that even more energy was emitted above  200 MeV. 
The third and the last burst to consider,  GRB 970417a was the burst (of 54 in the field of view) for which the Milagrito collaboration reported marginal evidence of TeV emission during the duration of the burst.  BATSE determined the burst position to be R.A. = 295.7$^\circ$ , decl. = 55.8$^\circ$  and its detection was  relatively weak with a fluence in the 50 - 300 keV energy range of $1.5\times10^{-7} ergs/cm^2$  and $T_{90}$ observed with BATSE of 7.9 s.

\section{External Shock Model}

We have considered a leptonic model, where electrons are accelerated in external shock (forward and reverse shocks). This external shock also generates magnetic fields \citep{med99}.  However, unlike the forward shock emission that continues later at lower energy, the reverse shock emits a single burst in the  $\gamma$- or X-ray band. The difference between our formalism and previous ones is that, we describe in a unified way ($\mathcal{R}_B, \mathcal{R}_e, \mathcal{R}_x, \mathcal{R}_M$, see \citep{fra12a})  the high-energy emission through a superposition of SSC processes from  forward and reverse shocks. We summarize the model below.  GRB emission is produced when an expanding relativistic shell interacts with  the  circumburst medium producing forward and reverse shocks.  In each shock the constant fractions $\epsilon_{e,f/r}$ and $\epsilon_{B,f/r}$ of the shock energy go into electrons and the magnetic field respectively. For the forward  shock,  we assume that electrons are accelerated to a power-law distribution of Lorentz factor $\gamma_e$ with a minimum Lorentz factor {\footnotesize $ \gamma_m: N(\gamma_e)\,d\gamma_e \propto \gamma_e^{-p}\,d\gamma_e$}  with {\footnotesize$ \gamma_e\geq\gamma_m $} and {\footnotesize$\gamma_m=\epsilon_{e,f}(p-2)/ (p-1) m_p/m_e\,\gamma_f$},  where  {\footnotesize $\epsilon_{B,f}=  B_f^2/(32\pi\,\gamma^2_f\,\eta_f\,m_p)$}   and {\footnotesize$\epsilon_{e,f}=  U_e/(4\,\gamma^2_f\,\eta_f\,m_p)$} are  the  magnetic and electron equipartition parameters respectively, $\gamma_f$ is the Lorentz factor of the bulk and $\eta_f$ is the ISM density.   Given the cooling electron Lorentz factor  {\footnotesize$ \gamma_{e,c}=3\,m_e(1+z)/(16\,\epsilon_{B,f}\,\sigma_T\,m_p\,t_{d,f}\,\Gamma_f^3\,\eta_f)$}  and the deceleration time $t_{d,f}$,  the break energies  of the photons radiated by electrons at a distance $D$ from the source in natural units (c=$\hbar$=1) are given by,

%\begin {small}
 \bary\label{synforw}\nonumber
E_{\rm m,f}&\sim& \frac{2^{5/2}\,\pi^{1/2}\, q_e\,m_p^{5/2}\, (p-2)^2}{m_e^3\,(p-1)^2} \, (1+z)^{-1}\,\epsilon_{e,f}^2\,\epsilon^{1/2}_{B,f}\,n^{1/2}_{f}\,\gamma^{4}_{f}\cr
E_{\rm c,f}&\sim& \frac{\pi^{7/6}\, 3^{4/3}\,m_eq_e}{2^{13/6}\,m_p^{5/6}\,\sigma^2_T}\, (1+z)^{-1}\,(1+x_f )^{-2}\,\epsilon^{-3/2}_{B,f}\,n^{-5/6}_{f}\,E^{-2/3}\,\gamma^{4/3}_{f}\cr
\eary
%\end{small}

\noindent where $E$ is the isotropic energy. The SSC break energies  ({\footnotesize $E^{(\rm IC)}_{m,f}\sim\gamma^2_{m},E_{m,f}$} and    {\footnotesize$E^{(IC)}_{c,f}\sim\gamma^2_c\,E_{c,f}$})   are also given by\cite{sac12},
%\begin {small}
\bary\label{sscf} \nonumber
E^{(IC)}_{\rm m,f}&\sim& \frac{6\,q_e\, m_p^{15/4}}{2^{5/4}\,(3\,\pi)^{1/4}\,m_e^5} \, (1+z)^{5/4}\,\epsilon_{e,f}^{4}\,\epsilon_{B,f}^{1/2}\,n^{-1/4}_{f}\,E^{3/4}\,t_{f}^{-9/4}\cr
E^{(IC)}_{\rm c,f}&\sim& \frac{2^{3/4}\,27\, \pi^{7/4}\,q_e\,m_e^3}{128\,3^{1/4}\,m_p^{9/4}\,\sigma_T^4 } \, (1+z)^{-3/4}\,(1+x_f )^{-4}\,\epsilon_{B,f}^{-7/2}\,n^{-9/4}_{f}\,E^{-5/4}\,t_{f}^{-1/4}\cr
\eary
%\end{small}
On the other hand, when  the  reverse shock crosses the shell it heats up and accelerates electrons. Considering the thick shell case, when the ejecta is significantly decelerated, the  synchrotron 
%\begin {small}
\bary\label{synrev}\nonumber
E_{\rm m,r}&\sim&  \frac{4\,\pi^{1/2}\,q_e\,m_p^{5/2}\,(p-2)^2}{m_e^3\,(p-1)^2}  \,(1+z)^{-1}\,\epsilon_{e,r}^{2}\,\epsilon_{B,r}^{1/2}\,\Gamma^{2}_{r}\,n^{1/2}_{r} \cr
E_{\rm c,r}&\sim&  \frac{9\pi\,2^{1/2}\,m_e\,q_e}{8\,(3^{1/2})\,m_p\,\sigma^2_T}   \,(1+z)^{-1/2}\,(1+x_r+x_r^2)^{-2}\,\epsilon_{B,r}^{-3/2}\,n^{-1}_{r}\,E^{-1/2}\,T_{90}^{-1/2}\cr
%\eary
%\end{small}
%\begin {small}
%\bary\label{ssc}\nonumber
\eary
and SSC break energies are given by \cite{fra12a,fra12b},

\bary\nonumber
E^{(IC)}_{\rm m,r}&\sim&  \frac{2^{21/4}\pi^{3/4}\,m_p^{13/4}\,(p-2)^4}{3^{1/4}\,m_e^5\,(p-1)^4} \,(1+z)^{-7/4}\,\epsilon_{e,r}^{4}\,\epsilon_{B,r}^{1/2}\,\Gamma^{4}_{r}\,n^{3/4}_{r}\,E^{-1/4}\,T_{90}^{3/4}\,,\cr
E^{(IC)}_{\rm c,r}&\sim&  \frac{3^{7/2}\pi\, m_e^3\,q_e}{2^{11}\,m_p^3\,\sigma_T^4   }\,(1+z)^{3/2}\,(1+x+x^2)^{-4}\,\epsilon_{B,r}^{-7/2}\,n^{-3}_{r}\,E^{-1/2}\,\Gamma^{-6}_{r}\,T_{90}^{-5/2}\,,\cr
\eary
%\end{small}
\noindent where $T_{90}$ is the burst duration. A detailed description of the model is given in Fraija et al. (2012) and Sacahui  et al. (2012).

\section{Results and Conclusions}
 
 We use typical \cite{fra12a,fra12b,fra12c} values  (table 1) for $\epsilon_{B,r}$ and $\epsilon_{B,f}\sim (10^{-4}-10^{-3})$ which  in comparison with previous works on GRB 980923 and GRB 090926A,  do not require to be different,  $\epsilon_{e,r}\neq\epsilon_{e,f}\sim (0.1-0.9)$. Also,  $\eta_{f}\leq\eta_{r} \sim 10\,\rm{cm^{-3}}$,  $\gamma_f\sim$ 600 and $\gamma_r\sim$ 1000. The calculated and observed quantities are given in Table 2.   In this letter,  we  present  a leptonic model based on external shocks to describe  the  high-energy emission  for GRB 940217, GRB 941017 and GRB 970217A.  Clearly the observations for the considered bursts are  less restrictive than those for GRB 980923 and GRB 090926A\cite{sac12,fra12a}.  To describe  the high energy component  in GRB 940217 and GRB 970417A was only required  SSC emission from forward shock and while for  GRB 941017,  the high  component $\geq$200 MeV  was  a superposition of SSC emission from forward and reverse shock in the thick shell case.  Unlike of GRB 980923 and GRB 090926A, was not  required the magnetization of the jet.  This may be a consequence of the lack of keV-MeV emission to constrain the emission from the reverse shock.  We note that there could be some bursts with emission at energies  up to a few TeV, which would be candidates to be detected by observatories with wide-field of view as HAWC\cite{aba12}.

\begin{center}\renewcommand{\arraystretch}{0.4}\addtolength{\tabcolsep}{-1pt}
\begin{tabular}{ l c c c }
  \hline
 {GRBs} & {940217} &  {941017}   & {970217A}  \\
 \hline
%{\footnotesize Input parameters}   &  &    &\\

\normalsize{Forward shock}  & & &  \\

%%%%%%%%%%%%%%%%%%%%%%%%%%%%%%%%%%%%%%%%%%%%%%%%%%%%%%%%%%%%%%%%%%%%%%%%%%%%%%%%%

{\footnotesize$\epsilon_{B,f}$} & {\footnotesize$10^{-3}$}&   {\footnotesize$10^{-3}$} &{\footnotesize$10^{-4}$}  \\
{\footnotesize$\epsilon_{e,f}$} &{\footnotesize$0.3$}& {\footnotesize$0.1$} &{\footnotesize$0.5$}  \\
{\footnotesize$n_f$ ($cm^{-3}$)} &  {\footnotesize 1} &  {\footnotesize 10} & {\footnotesize 1}  \\
{\footnotesize$\gamma_f$} & {\footnotesize 600} & {\footnotesize 600} &{\footnotesize 600}   \\
\hline

\normalsize{Reverse shock}  & & &  \\

%%%%%%%%%%%%%%%%%%%%%%%%%%%%%%%%%%%%%%%%%%%%%%%%%%%%%%%%%%%%%%%%%%%%%%%%%%%%%%%%%

%{\footnotesize Input parameters}   &  &    &\\
{\footnotesize$\epsilon_{B,r}$} & {\footnotesize$10^{-3}$}&   {\footnotesize$10^{-3}$} &{\footnotesize$10^{-4}$}  \\
{\footnotesize$\epsilon_{e,r}$} &{\footnotesize$0.5$}& {\footnotesize$0.9$} &{\footnotesize$0.5$}  \\
{\footnotesize$n_r$ ($cm^{-3}$)} &  {\footnotesize 10} &  {\footnotesize 10} & {\footnotesize 10}  \\
{\footnotesize$\gamma_r$} & {\footnotesize 1000} & {\footnotesize 1000} &{\footnotesize 600}   \\
\hline

%\normalsize{REVERSE SHOCK} & &   \\

%%%%%%%%%%%%%%%%%%%%%%%%%%%%%%%%%%%%%%%%%%%%%%%%%%%%%%%%%%%%%%%%%%%%%%%%%%%%%%%%%

%\normalsize{Input parameters} & &  \\
%\scriptsize{$\epsilon_{B,r}$} &  \scriptsize{-} & \scriptsize{-}   \\
%\scriptsize{$\epsilon_{e,r}$} & \scriptsize{-} & \scriptsize{-} \\
%\scriptsize{$n_r$ ($cm^{-3}$)} & \scriptsize{-} & \scriptsize{-} \\
%\scriptsize{$\gamma_r$} &\scriptsize{-} & \scriptsize{-} \\
%%%%%%%%%%%%%%%%%%%%%%%%%%%%%%%%%%%%%%%%%%%%%%%%%%%%%%%%%%%%%%%%%%%%%%%%%%%%%%%%%
 %\hline
 
\end{tabular}
\end{center}

\begin{center}
{\footnotesize \textbf{Table 1. Parameters used.}}\\
\end{center}

\begin{center}\renewcommand{\arraystretch}{0.5}\addtolength{\tabcolsep}{-5.0pt}
\begin{tabular}{ l c c c c c}
  \hline
 {GRBs} &{940217} & {941017} & {970417A}   \\
 \hline
{\footnotesize Quantities} & { \footnotesize calculated (observed)}       & {\footnotesize calculated (observed)} &  {\footnotesize calculated (observed)} \\
\normalsize{Forward shock} & &  \\

%%%%%%%%%%%%%%%%%%%%%%%%%%%%%%%%%%%%%%%%%%%%%%%%%%%%%%%%%%%%%%%%%%%%%%%%%%%%%%%%%

%\normalsize{Input parameters} &  &  &  &  \\
%\scriptsize{$\epsilon_{B,f}$} & \scriptsize{$10^{-4.3}$}  & \scriptsize{$10^{-4}$} & \scriptsize{$10^{-4.3}$}    \\
%\scriptsize{ $\epsilon_{e,f}$} & \scriptsize{$10^{-1}$}  & \scriptsize{0.17} & \scriptsize{$10^{-1}$}   \\
%\scriptsize{$n_f$ ($cm^{-3}$)} & \scriptsize{10}  & \scriptsize{10} & \scriptsize{1}   \\
%\scriptsize{$\gamma_f$} & \scriptsize{600}  & \scriptsize{900} & \scriptsize{600}   \\

%%%%%%%%%%%%%%%%%%%%%%%%%%%%%%%%%%%%%%%%%%%%%%%%%%%%%%%%%%%%%%%%%%%%%%%%%%%%%%%%%

 {\footnotesize $E_{\rm m,f}$ (keV)} &{\footnotesize 128.9 ( -)}& {\footnotesize 45.3 ( $\sim$ 100 )}    & {\footnotesize 169.8 ( $\sim$ 100 )}   \\
 {\footnotesize $E_{\rm c,f}$(eV)} &{\footnotesize 0.12 ( - )}&  {\footnotesize 111.3 ( - )}    & {\footnotesize 512 ( - )}  \\
%\scriptsize{$(\nu F_{\rm \nu max})^{syn}\,(erg\,cm^{-2}\,s^{-1})$} & \scriptsize{$1.41 \times 10^{-7}$}  & \scriptsize{$8.14 \times 10^{-7}$} & \scriptsize{$5.21 \times 10^{-8}$}   \\
{\footnotesize$E^{(IC)}_{\rm m,f}$(GeV)} &{\footnotesize33.13 ( $\sim 10$ )}  & {\footnotesize1.8 ( $\geq$0.2 )}   & { \footnotesize 8.6$\times10^3$ (1)}  \\
 {\footnotesize$E^{(IC)}_{\rm c,f}(eV)$} &  {\footnotesize$  3.8\times10^{-11}$ ( - )} & {\footnotesize$1.5 \times10^{3}$ ( - )}    & {\footnotesize885.2 ( - )}  \\
{\footnotesize Duration of the component (s)} &{\footnotesize1000 ( $\sim 5600$ ) }  & {\footnotesize150 ( $\geq 120$ ) }  &  {\footnotesize100 ( $\sim 100$ ) }    \\  

 {\footnotesize$(\nu F_{\rm \nu max})^{SSC}\,(erg\,cm^{-2}\,s^{-1})$} & {\footnotesize$2.95\times10^{-7}$  ( $\sim 10^{-7}$ )}  & {\footnotesize$3.74\times10^{-6}$  ( $\sim 10^{-6}$ )} & {\footnotesize $1.85\times10^{-5}$ ( $\sim 10^{-5}$ )}  \\
 
%%%%%%%%%%%%%%%%%%%%%%%%%%%%%%%%%%%%%%%%%%%%%%%%%%%%%%%%%%%%%%%%%%%%%%%%%%%%%%%%%
%%%%%%%%%%%%%%%%%%%%%%%%%%%%%%%%%%%%%%%%%%%%%%%%%%%%%%%%%%%%%%%%%%%%%%%%%%%%%%%%%
%%%%%%%%%%%%%%%%%%%%%%%%%%%%%%%%%%%%%%%%%%%%%%%%%%%%%%%%%%%%%%%%%%%%%%%%%%%%%%%%%
 \hline
\normalsize{Reverse shock} & &  \\

%%%%%%%%%%%%%%%%%%%%%%%%%%%%%%%%%%%%%%%%%%%%%%%%%%%%%%%%%%%%%%%%%%%%%%%%%%%%%%%%%

%\normalsize{Input parameters} & &  \\
%\scriptsize{$\epsilon_{B,r}$} & \scriptsize{0.125}  & \scriptsize{0.125} & \scriptsize{0.125}  \\
%\scriptsize{$\epsilon_{e,r}$} & \scriptsize{0.65}  & \scriptsize{0.87} & \scriptsize{0.65} \\
%\scriptsize{$n_r$ ($cm^{-3}$)} &\scriptsize{10}  & \scriptsize{100} & \scriptsize{10} \\
%\scriptsize{$\gamma_r$} & \scriptsize{1000}  & \scriptsize{1800} & \scriptsize{1000} \\
%%%%%%%%%%%%%%%%%%%%%%%%%%%%%%%%%%%%%%%%%%%%%%%%%%%%%%%%%%%%%%%%%%%%%%%%%%%%%%%%%

%{\footnotesize Quantities} & {\footnotesize calculated (observed)}       & {\footnotesize calculated (observed)} &  {\footnotesize calculated (observed)}  \\
 {\footnotesize$E_{\rm m,r}$ (eV)} & {\footnotesize 0.09 ( -  )}  &{\footnotesize 28.3  ( -  )}  & {\footnotesize 8.7  ( - )} \\
 {\footnotesize$E_{\rm c,r}$(eV)} & {\footnotesize 1.5 ( - )}  & {\footnotesize 9.8 ( - )}  & {\footnotesize 2.93 ( - )} \\
%\scriptsize{$(\nu F_{\rm \nu max})^{syn}\,(erg\,cm^{-2}\,s^{-1})$} & \scriptsize{$3.9 \times 10^{-6}$}  & \scriptsize{$5.4 \times 10^{-7}$} & \scriptsize{$3.7 \times 10^{-7}$ } \\
 {\footnotesize$E^{(IC)}_{\rm m,r}$(MeV)} & {\footnotesize $1.3\times10^{-3}$ ( - ) }  &{\footnotesize  386.5 ( $\geq$200 ) }  & {\footnotesize 13.1 ( -)}  \\
 {\footnotesize$E^{(IC)}_{\rm c,r}(eV)$} &  {\footnotesize 152.9 ( - )}  &{\footnotesize 386.5 ( - )}  & {\footnotesize 849.2 ( - )}   \\
 {\footnotesize $(\nu F_{\rm \nu max})^{SSC}\,(erg\,cm^{-2}\,s^{-1})$} & {\footnotesize  $1.8\times10^{-7}$ ( - )} &{\footnotesize    $1.58\times10^{-6}$  ( $\sim 10^{-6}$ )}  & {\footnotesize   $8.8\times10^{-6}$ ( -)}   \\
 \hline

% \scriptsize{$1.2\times10^{-4}$} & \scriptsize{$1.33\times10^{-4}$}&\scriptsize{$1.04\times10^{-4}$} 

\end{tabular}
\end{center}

\begin{center}
{\footnotesize \textbf{Table 1. Calculated quantities using the model described in the text. When available, the observed values are given. }}\\
\end{center}

%\begin{theacknowledgments}
This work is partially supported by DGAPA-UNAM (Mexico) Project 
No. IN101409  and Conacyt Project No. 105033. 

%\end{theacknowledgments}



\bibliographystyle{aipproc}   % if natbib is available

\bibliography{sample}

\IfFileExists{\jobname.bbl}{}
 {\typeout{}
  \typeout{******************************************}
  \typeout{** Please run "bibtex \jobname" to optain}
  \typeout{** the bibliography and then re-run LaTeX}
  \typeout{** twice to fix the references!}
  \typeout{******************************************}
  \typeout{}
 }

%\end{document}

%%%%%%%%%%%%%%%%%%%%%%%%%%%%%%%%%%%%%%%%%%%
%% The following lines show an example how to produce a bibliography
%% without the help of the BibTeX program. This could be used instead
%% of the above.
%%%%%%%%%%%%%%%%%%%%%%%%%%%%%%%%%%%%%%%%%%%

\end{document}